\documentclass[a4paper,10pt]{article}
\usepackage{amssymb}
\usepackage{graphicx}
\usepackage{amsmath}
\usepackage{indentfirst}
\usepackage{mathrsfs}

\begin{document}


\title{Renormalizability of Generalized Quantum Electrodynamics}

\author{R. Bufalo$^{1}$\thanks{%
rbufalo@ift.unesp.br}~, B.M. Pimentel$^{1}$\thanks{%
pimentel@ift.unesp.br}~ and G. E. R. Zambrano$^{2}$\thanks{%
gramos@udenar.edu.co}~ \\
\textit{{$^{1}${\small Instituto de F\'{\i}sica Te\'orica (IFT/UNESP), UNESP
- S\~ao Paulo State University}}} \\
\textit{\small Rua Dr. Bento Teobaldo Ferraz 271, Bloco II Barra Funda, CEP
01140-070 S\~ao Paulo, SP, Brazil}\\
\textit{{$^{2}${\small Departamento de F\'{\i}sica, Universidad de Nari\~{n}o%
}}} \\
\textit{{\small Clle 18 Cra 50, San Juan de Pasto, Nari\~{n}o, Colombia}}}
\date{}
\maketitle

\begin{abstract}
In this work we present the study of the renormalizability of the
Generalized Quantum Electrodynamics ($GQED_{4}$). We begin the article by
reviewing the on-shell renormalization scheme applied to $GQED_{4}$.
Thereafter, we calculate the explicit expressions for all the counter-terms
at one-loop approximation and discuss the infrared behavior of the theory as
well. Next, we explore some properties of the effective coupling of
the theory which would give an indictment of the validity regime of theory: $m^{2} \leq k^{2} < m_{P}^{2}$. Afterwards, we make use of experimental data from the electron anomalous magnetic moment to set possible values for the theory
free parameter through the one-loop contribution of Podolsky mass-dependent term to Pauli's form factor $F_{2}\left(q^{2}\right)$.
\end{abstract}

\maketitle
\section{Introduction}

Effective theories may or may not be unitary. In fact, the unitarity is lost when a particle, retained in an effective theory, can lower its energy by the emission of other particles which have been eliminated in deriving the effective theory \cite{1}. The imprint of effective theories is the appearance of higher order derivatives terms in the Lagrangian density, reflecting the momentum-dependence of self-energies or form factors. Once the effective theory is rendered ultraviolet (UV) finite, we may consider it as an extension of the class of potentially interesting and consistent models because its UV dynamics is well defined. Motivated by the search of possible fundamental theories, one naturally expects the complete suppression of nonphysical, nonunitary processes.

Higher derivative (HD) theories \cite{2} were proposed in an attempt to tame the ultraviolet behavior of physically relevant models \cite{3}. However, it was soon recognized that they have a Hamiltonian which is not bounded from below \cite{4} and that the addition of such terms leads to the existence of negative norm states (or ghosts states) -- induces an indefinite metric in the space of states -- jeopardizing thus the unitarity \cite{5}. Despite the fact that many attempts to overcome these ghost states have been proposed, no one has been able to give a general method to deal with them \cite{27,28,26}. In fact, in conventional gauge theories the gauge-fixing term, the Faddeev-Popov-De Witt ghosts, and the original Lagrangian density are invariant under BRST symmetry. Its purpose is to remove all the negative norm states and get a unitary theory, physical states are thus defined as those which have zero ghost number. In the light of this idea it was proposed that the HD theories present a BRST symmetry, which is seen as an intrinsic feature of these HD theories \cite{13}. However, by imposing this symmetry we were led to an unitary but trivial resulting theory through the quartet mechanism \cite{21}, since all physical states, excepted the vacuum, appear in zero-norm combinations.

It is known that HD theories have, as a field theory, better renormalization properties than the conventional ones. This idea appeared to be quite successful in the case of the attempt to quantize gravity, where the Einstein action is supplied by terms containing higher powers of curvature leading to a renormalizable \cite{29} and asymptotically free theory \cite{27}. Also, a new impetus in exploring appealing quantum theories such as $f(R)$-gravity \cite{25}, which may explain the accelerating universe.

One of the first HD theories was the Generalized Electrodynamics \cite{6} originally proposed as to get
rid of some pathologies inherent in the Maxwell theory. As pointed out in
several works \cite{6,7}, it is clear that the Maxwell's theory is not the
only one to describe the Electromagnetic Field and therefore considered as an appealing theory. In fact, the Podolsky's theory is the only possible linear, Lorentz and $U(1)$ invariant generalization of Maxwell's theory \cite{8}. Actually, it is impressive that the simplest $U(1)$ generalization possesses a richness of interesting features. Concerning the gauge symmetry, it was proved in Ref. \cite{9} that the generalized Lorenz
condition, $\Omega \left[ A\right] =\left( 1+m_{P}^{-2}\square \right)
\partial ^{\mu }A_{\mu }$, is the correct gauge condition which
completely fix the gauge freedom for the Podolsky's theory. Such generalized condition
is attached with the correct true degrees of freedom for the theory. A
recent study of the finite-temperature free Podolsky's theory has showed a correction to the Stefan-Boltzmann law and by using cosmic microwave
background data it was possible to set a thermodynamical limit to the
Podolsky's parameter $m_{P}$ \cite{10}.

The issue of quantization of Generalized Electrodynamics in the generalized Lorenz condition were dealt in Ref. \cite{11} through functional methods. In contrast to the results obtained in the usual Lorenz condition \cite{12} it was found that, in the one-loop approximation, the electron self-energy and the vertex function are both ultraviolet finite. Therefore, a natural continuation of these studies would be a discussion upon the renormalizability of the Generalized Quantum Electrodynamics ($GQED_{4}$). Efforts have also been made to analyze
the $GQED_{4}$ at thermodynamic equilibrium \cite{14}; so far, the
current analysis lies upon formal aspects of the theory.

It is worth noticing some points about the free gauge field Green's function \cite{11}:
\begin{eqnarray}
iD_{\mu \nu }\left( k\right)  &=&\frac{1}{k^{2}}\left[ \eta _{\mu \nu
}-\left( 1-\xi \right) \frac{k_{\mu }k_{\nu }}{k^{2}}\right] -\frac{1}{k^{2}-m_{P}^{2}}\left[ \eta _{\mu \nu }+\left( 1-\xi \right) \frac{k_{\mu
}k_{\nu }}{k^{2}-m_{P}^{2}}\right]   \label{eq 2.1} \\
&&+\left( 1-2\xi \right) \frac{1}{\left( k^{2}-m_{P}^{2}\right) k^{2}}k_{\mu
}k_{\nu }+\frac{1}{\left( k^{2}-m_{P}^{2}\right) ^{2}}k_{\mu }k_{\nu }. \notag
\end{eqnarray}%
The above expression shows explicitly the contributions of the ghost states -- even the Maxwell's theory is correctly quantized only in an indefinite-metric space \cite{30}, but they can safely be excluded from the
asymptotic states by means of gauge symmetry without upsetting the unitary time evolution in the physical, positive
norm sector. Actually there are sufficient reasons to dismiss any quantum field theory, such as Einstein gravity, that had the presence of HD quantum corrections and ghosts states; nevertheless, in this paper we will assume that, although there is not a formal proof upon the details of physical space for Podolsky's theory, it could be performed an analysis, for instance through either the generalized Kubo-Martin-Schwinger boundary conditions \cite{31}, through BRST symmetry and quartet mechanism \cite{13}, or even by the quantization scheme proposed by Hawking and Hertog \cite{28}, leading therefore to an acceptable and well-defined theory to the Generalized Electrodynamics.\footnote{The arguments upon unitarity and consistency of HD theories presented on these works can be suitably extended for the Generalized Electrodynamics.}

This work is addressed to the issue of renormalizability of the Generalized
Quantum Electrodynamics, the plan of the paper is as follows: In Sec.\ref%
{sec:1}, we review and apply to the $GQED_{4}$ the on-shell renormalization
prescription with the appropriated renormalization
conditions. In Sec.\ref{sec:2}, we compute the expressions for the
counter-terms, and the infrared behavior of the theory is also taken into
account. In Sec.\ref{sec:3}, we present a proper discussion on the
effective coupling of the theory. In Sec.\ref{sec:4}, we calculate, at
one-loop approximation, the Podolsky's contribution to the electron
anomalous magnetic moment and accordingly to experimental data we set a
value to the theory's free parameter $m_{P}$. We present our remarks and
prospects in the Sec.\ref{sec:5}. The Minkowski spacetime is concerned in
the whole work, with the metric signature $(+,-,-,-)$.

\section{Renormalization Schedule}

\label{sec:1}

As noticed before, the $GQED_{4}$ shows a better UV behavior than the usual theory \cite{11}, however it contains negative norm states, which induce the lost of unitarity. This is a major issue and certainly cannot be neglected, and requires a detailed study; nevertheless, once we have identified the primitive divergences on the $1PI$ functions \cite{11} and the theory rendered UV finite (at the light of effective theories) we are motivated to retain some attention regarding the underlying aspects of the renormalized behavior of the present theory. We expect to obtain from comparison of our results to the regular expressions for the effective coupling and one-loop contribution for the anomalous magnetic moment, an energy regime of validity of the theory and a bound value for the free parameter, respectively.

In the following, we shall give a brief derivation on the so-called on-shell renormalization scheme
\cite{15} by employing it into the $GQED_{4}$. Although the renormalization procedure be a well-known subject in the quantum field theory, we would like to point out some relevant features in which the discussion for $GQED_{4}$ may differ from the usual one. The
first part of the analysis relies by determining formally the
constants $Z_{i}$ under suitable renormalization conditions. Therefore, to this aim, we define the
gauge-fixed renormalized Lagrangian, in the gauge choice of
the so-called generalized Lorenz condition \cite{9}: $\Omega \left[ A\right] =\left(
1+m_{P}^{-2}\square \right) \partial ^{\mu }A_{\mu }$, and also
introduce the counter-terms as the following prescription:
\begin{eqnarray}
\mathcal{L} &=&\bar{\psi}\left( i\hat{\partial}-m+e\hat{A}\right) \psi -%
\frac{1}{4}F_{\mu \nu }F^{\mu \nu }  +\frac{1}{2m_{P}^{2}}\partial ^{\mu
}F_{\mu \beta }\partial _{\alpha }F^{\alpha \beta }-\frac{1}{2\xi }\left[
\left( 1+m_{P}^{-2}\square \right) \partial ^{\mu }A_{\mu }\right] ^{2}
\label{eq 1.5} \\
&&+\delta _{Z_{2}}\bar{\psi}i\hat{\partial}\psi -\delta _{Z_{0}}\bar{\psi}%
m\psi +\delta _{Z_{1}}e\bar{\psi}\hat{A}\psi -\delta _{Z_{3}}\frac{1}{4}%
F_{\mu \nu }F^{\mu \nu }; \notag
\end{eqnarray}%
with the following definition: $\delta _{Z_{i}}=Z_{i}-1$. The relations
between the bare and renormalized quantities are, as usual, as follows:%
\begin{equation}
A^{\left( 0\right) }=Z_{3}^{1/2}A^{\left( r\right) },~~ \psi ^{\left(
0\right) }=Z_{2}^{1/2}\psi ^{\left( r\right) },~~ \bar{\psi}^{\left(
0\right) }=Z_{2}^{1/2}\bar{\psi}^{\left( r\right) },  \label{eq 1.2a}
\end{equation}%
and\footnote{It could also be introduced: $m_{0}=Z_{m}m$, with $Z_{m}=\frac{Z_{0}}{Z_{2}}$.}
\begin{equation}
Z_{2}m^{\left( 0\right) }=Z_{0}m,~~ Z_{3}^{1/2}e^{\left( 0\right)
}=Z_{1}Z_{2}^{-1}e,~~ m_{P}^{\left( 0\right) }=Z_{3}^{1/2}m_{P}.
\label{eq 1.2}
\end{equation}%
Here, the Podolsky's parameter $m_{P}$ has not a constant associated with its
renormalization, in the same sense as the $\xi$ parameter (gauge Ward-Fradkin-Takahashi identity \cite{11}).

As in the usual $QED_{4}$, we obtain here from the bare Ward-Fradkin-Takahashi identity \cite{11}:
\begin{equation}
ik_{\mu }\tilde{\Gamma}^{\mu }\left( p,p^{\prime };q=p^{\prime }-p\right) =%
\mathscr{S}^{-1}\left( p-p^{\prime }\right) -\mathscr{S}^{-1}\left( p\right),
\end{equation}%
that the ratio $Z_{1}/Z_{2}$ must be finite if the theory is renormalizable. Thus, the finiteness of the ratio $Z_{1}/Z_{2}$ implies that
order-by-order in perturbation theory the equality $Z_{1}=Z_{2}$ is
identically satisfied. Thereby, the coupling constant $e$ is determined by $Z_{3}$ only: $%
e_{0}=Z_{3}^{-1/2}e$.

According to the Lagrangian \eqref{eq 1.5} we obtain here
new Schwinger-Dyson-Fradkin equations for the theory, which we shall denote by the suffix $^{\left( R\right) }$. More precisely, the original
self-energy functions \cite{11} are changed by adding the counter-terms $\delta
_{Z_{i}}$. Now, by a formal derivation it is not complicated to show that the photon sector has the following renormalized self-energy function:%
\begin{equation}
\Pi ^{\left( R\right) }\left( k\right) =\Pi \left( k\right) +\delta _{Z_{3}};
\label{eq 1.7}
\end{equation}%
where $\Pi \left( k\right) $ is the polarization scalar written in terms of
the renormalized quantities \cite{11}.

We proceed now to the first renormalization condition, which comes to ensure the Maxwell photon behavior for the propagator Eq.\eqref{eq 2.1}, in the gauge $ \xi =1$. Thus, it is written as the following:
\begin{equation}
i\mathscr{D}_{\mu \nu }\left( k\right) =\frac{1}{k^{2}}\eta _{\mu \nu
},\quad \text{when }k^{2}\rightarrow 0.  \label{eq 1.8}
\end{equation}%
By means of the above condition we find the expression for the counter-term $\delta
_{Z_{3}} $:%
\begin{equation}
\delta _{Z_{3}}=Z_{3}-1=-\left. \Pi \left( k^{2}\right) \right\vert
_{k^{2}\rightarrow 0}.  \label{eq 1.10}
\end{equation}
Considering next the fermionic sector, we have that
its renormalized self-energy function is written as:%
\begin{equation}
i\Sigma ^{\left( R\right) }\left( p,m\right) =i\Sigma \left( p,m\right)
-im\delta _{Z_{0}}+i\delta _{Z_{2}}\widehat{p};  \label{eq 1.11}
\end{equation}%
where the function $\Sigma \left( p\right) $ is the radiative correction of
the fermionic $1PI$ function: $\Gamma \left( p\right) =\widehat{p}-m-\Sigma
^{\left( R\right)}\left( p,m\right) $; where: $\Gamma \left( x,y\right) =-%
\frac{\delta ^{2}\Gamma }{\delta \psi \left( y\right) \delta \bar{\psi}%
\left( x\right) }$. As a matter of calculation, we can write the electron
self-energy function in the following general way: $\Sigma \left( p,m\right)
=\Sigma _{1}\left( p^{2}\right) \widehat{p}+\Sigma _{2}\left( p^{2}\right)
I$. In order to fix the fermionic counter-terms we must impose two
renormalization conditions. Which are outlined by:\footnote{$%
m_{F} $ is defined as the zero of the electron $1PI$ function.}%
\begin{equation}
\frac{\partial \Gamma \left( p\right) }{\partial \hat{p}}=1,~~ \Gamma \left( p\right) =\widehat{p}-m_{F},\quad \text{when }\widehat{p}%
\rightarrow m_{F}. \label{eq 1.13}
\end{equation}
These immediately lead to the following relations for the counter-term $\delta _{Z_{2}}$:
\begin{eqnarray}
\delta _{Z_{2}}=Z_{2}-1=-\left. \Sigma _{1}\left( p^{2}\right) \right\vert
_{p^{2}\rightarrow m_{F}^{2}}-2m_{F}^{2}\left. \frac{\partial \Sigma
_{1}\left( p^{2}\right) }{\partial p^{2}}\right\vert _{p^{2}\rightarrow
m_{F}^{2}}-2m_{F}\left. \frac{\partial \Sigma _{2}\left( p^{2}\right) }{%
\partial p^{2}}\right\vert _{p^{2}\rightarrow m_{F}^{2}},   \label{eq 1.13c}
\end{eqnarray}%
whereas for the counter-term $\delta _{Z_{0}}$ one gets:
\begin{eqnarray}
m\delta _{Z_{0}}=m\left( Z_{0}-1\right) =\left. \Sigma _{2}\left(
p^{2}\right) \right\vert _{p^{2}\rightarrow m_{F}^{2}} -2m_{F}^{2}\left[m_{F}\left. \frac{\partial \Sigma _{1}\left( p^{2}\right) }{\partial p^{2}}%
\right\vert _{p^{2}\rightarrow m_{F}^{2}}+\left. \frac{\partial \Sigma
_{2}\left( p^{2}\right) }{\partial p^{2}}\right\vert _{p^{2}\rightarrow
m_{F}^{2}}\right] .  \label{eq 1.16}
\end{eqnarray}%

Finally, we look on the renormalization condition which determines the counter-term $\delta _{Z_{1}}$. Using the
so-called Gordon decomposition, we can write the vertex part $\Lambda $ in
terms of the Dirac and Pauli form factors, as:%
\begin{equation}
\Lambda ^{\rho }\left( p,p^{\prime }\right) =\gamma ^{\rho }F_{1}\left(
q^{2}\right) +\frac{i}{2m}\sigma ^{\rho \nu }q_{\nu }F_{2}\left(
q^{2}\right) ,  \label{eq 1.17}
\end{equation}%
where $q=p^{\prime }-p$, is the transferred momentum and $\sigma ^{\rho \nu }=\frac{i}{2}\left[ \gamma ^{\rho
},\gamma ^{\nu }\right]$. Therefore, as a result of the on-shell condition for the vertex part: $p^{\prime 2}=p^{2}=m^{2}$, and $q^{2}\rightarrow 0$, we have:%
\begin{equation}
\left. F_{1}\left( q^{2}\right) \right\vert _{q^{2}\rightarrow 0}=0,
\label{eq 1.18}
\end{equation}%
which results in determining the counter-term $\delta _{Z_{1}}$.

The above discussion was in fact necessary to elaborate some details on the derivation of the formal expressions
for the counter-terms $\delta _{Z_{i}}$, to thus we become clear on some points of the renormalized structure of the present theory.\footnote{Technically it is not complicated to see through a perturbative graph analysis that the counter-terms are sufficient to absorb all the primitive divergences order-by-order.} In the next section we shall present the one-loop expressions for the three self-energy functions at a general-$\xi$,
and then calculate the explicit expressions for the counter-terms $\delta_{Z_{i}}$. Afterwards, we discuss the infrared behavior of these quantities.


\section{Renormalization Constants and Infrared Behavior}

\label{sec:2}

We begin this section by presenting the explicit expressions for the
one-loop radiative corrections functions at a general $\xi$ \cite{11}. The expression for the polarization scalar reads as \cite{15}:
\begin{eqnarray}
\Pi ^{\left( 1\right) }\left( k\right) =-\frac{\alpha }{3\pi }\bigg[ \frac{2%
}{\epsilon _{UV}}-\frac{1+2\gamma }{2}+6\int_{0}^{1}dyy\left( 1-y\right) \ln %
\left[ \frac{4\pi \mu ^{2}}{m^{2}-y\left( 1-y\right) k^{2}}\right] \bigg] .\label{eq 2.2}
\end{eqnarray}%
Where $\epsilon _{UV}=4-d$, $\epsilon _{UV}\rightarrow 0^{+}$, is the
ultraviolet dimensional regularization parameter and $\mu$ the t'Hooft mass, and $\alpha =\frac{e^{2}}{%
4\pi }$, is the fine-structure constant. The photon sector will be further
investigate in the section \ref{sec:3}, on the discussion about the theory's effective coupling.

Unlike $\Pi $, the electron self-energy and vertex part are both
gauge-dependent. Thus, at one-loop calculation, the electron self-energy
expression can be casted into the following form \cite{11}:{\small
\begin{eqnarray}
&&\Sigma ^{\left( 1,\xi \right) }\left( p\right) =\frac{\alpha }{2\pi }%
\int_{0}^{1}dz\left( \widehat{p}\left( 1-z\right) -2m\right) \ln \left\vert
\frac{z-\left( 1-z\right) z\frac{p^{2}}{m^{2}}+\left( 1-z\right) \frac{m^{2}_{P}}{m^{2}}}{%
z-\left( 1-z\right) z\frac{p^{2}}{m^{2}}}\right\vert   \label{eq 2.3} \\
&&+\frac{\alpha }{4\pi }\left( \frac{p^{2}}{m^{2}}\right)
\int_{0}^{1}dx\int_{0}^{1-x}dy\left( \left( 1-y\right) \widehat{p}+m\right)
y^{2}\Bigg\{ \frac{\left( \xi -1\right)}{y-\left( 1-y\right) y\frac{p^{2}}{%
m^{2}}}+ \frac{\left( 1-2\xi \right)}{y-\left( 1-y\right) y\frac{p^{2}}{%
m^{2}}+x\frac{m_{P}^{2}}{m^{2}}} \notag \\
&&+ \frac{\xi}{y-\left( 1-y\right) y\frac{p^{2}}{m^{2}}+\left( 1-y\right)
\frac{m_{P}^{2}}{m^{2}}}\Bigg\}  \notag \\
&&+\frac{\alpha }{4\pi }\int_{0}^{1}dx\int_{0}^{1-x}dy\left( 2m-\left(
1+3y\right) \widehat{p}\right) \Bigg\{\left( \xi -1\right) \ln \left\vert
\frac{y-\left( 1-y\right) y\frac{p^{2}}{m^{2}}}{y-\left( 1-y\right) y\frac{%
p^{2}}{m^{2}}+x\frac{m_{P}^{2}}{m^{2}}}\right\vert+ \xi \ln \left\vert \frac{%
y-\left( 1-y\right) y\frac{p^{2}}{m^{2}}+\left( 1-y\right) \frac{m_{P}^{2}}{%
m^{2}}}{y-\left( 1-y\right) y\frac{p^{2}}{m^{2}}+x\frac{m_{P}^{2}}{m^{2}}}%
\right\vert \Bigg\}.  \notag
\end{eqnarray}%
}The form factors explicit expressions of the vertex part $\Lambda $ are given by \cite{11}:{\small%
\begin{eqnarray}
&&F_{1}\left( q^{2}\right) =\left( Z_{1}-1\right) ^{\left( 1\right) }+\frac{%
\alpha }{2\pi }\int_{0}^{1}dx\int_{0}^{1-x}dy\ln \left\vert \frac{%
(x+y)^{2}+\left( 1-x-y\right) \frac{m_{P}^{2}}{m^{2}}-xy\frac{q^{2}}{m^{2}}}{%
(x+y)^{2}-xy\frac{q^{2}}{m^{2}}}\right\vert \notag \\
&&+\frac{3\alpha }{2\pi }\int_{0}^{1}dx\int_{0}^{1-x}dy\int_{0}^{1-x-y}dz%
\left[ \left( \xi -1\right) \ln \left\vert \frac{(x+y)^{2}+z\frac{m_{P}^{2}}{%
m^{2}}-\frac{q^{2}}{m^{2}}xy}{(x+y)^{2}-xy\frac{q^{2}}{m^{2}}}\right\vert
+\xi \ln \left\vert \frac{(x+y)^{2}+z\frac{m_{P}^{2}}{m^{2}}-xy\frac{q^{2}}{%
m^{2}}}{(x+y)^{2}+\left( 1-x-y\right) \frac{m_{P}^{2}}{m^{2}}-xy\frac{q^{2}}{%
m^{2}}}\right\vert \right]  \notag\\
&&+\frac{\alpha }{4\pi }\int_{0}^{1}dx\int_{0}^{1-x}dy\left( 2\left(
(x+y)(x+5y)-2(1-x-y)^{2}-4x\right) +2\frac{q^{2}}{m^{2}}\left[ 1-x-y+xy%
\right] \right)\notag  \\
&&\times \left\{ \frac{1}{(x+y)^{2}-xy\frac{q^{2}}{m^{2}}}-\frac{1}{%
(x+y)^{2}+\left( 1-x-y\right) \frac{m_{P}^{2}}{m^{2}}-xy\frac{q^{2}}{m^{2}}}%
\right\} \notag \\
&&+\frac{\alpha }{2\pi }\int_{0}^{1}dx\int_{0}^{1-x}dy\int_{0}^{1-x-y}dz%
\left( 2(x+y)(2-x-y)-(1-x-y)^{2}+\frac{q^{2}}{2m^{2}}\left(
1-3x-3y+6xy\right) \right)   \notag \\
&&\times \Bigg\{\frac{\left( \xi -1\right) }{(x+y)^{2}-xy\frac{q^{2}}{m^{2}}}%
+\frac{\left( 1-2\xi \right) }{(x+y)^{2}+z\frac{m_{P}^{2}}{m^{2}}-xy\frac{%
q^{2}}{m^{2}}}+\frac{\xi }{(x+y)^{2}+\left( 1-x-y\right) \frac{m_{P}^{2}}{%
m^{2}}-xy\frac{q^{2}}{m^{2}}}\Bigg\}  \notag \\
&&+\frac{\alpha }{4\pi }\int_{0}^{1}dx\int_{0}^{1-x}dy\int_{0}^{1-x-y}dz\Big(%
\left( 4(x-xy+y)((x+y)(1-y)-x^{2})+(x^{2}+y^{2})^{2}\right)   \notag \\
&&-\frac{q^{2}}{m^{2}}\left( x(1-y)+y(1-x)\right) \left( 1-\left(
1-x-y\right) ^{2}\right) +2\left( \frac{q^{2}}{m^{2}}\right) ^{2}xy\left(
1-x\right) \left( 1-y\right) \Big)  \notag \\
&&\times \Bigg\{\frac{\left( \xi -1\right) }{\left[ (x+y)^{2}-xy\frac{q^{2}}{%
m^{2}}\right] ^{2}}+\frac{\left( 1-2\xi \right) }{\left[ (x+y)^{2}+z\frac{%
m_{P}^{2}}{m^{2}}-xy\frac{q^{2}}{m^{2}}\right] ^{2}}+\frac{\xi }{\left[
(x+y)^{2}+\left( 1-x-y\right) \frac{m_{P}^{2}}{m^{2}}-xy\frac{q^{2}}{m^{2}}%
\right] ^{2}}\Bigg\},  \label{eq 2.4}
\end{eqnarray}%
}and{\small
\begin{equation}
F_{2}\left( q^{2}\right) =\frac{2\alpha }{\pi }m^{2}\int_{0}^{1}dx%
\int_{0}^{1-x}dy\left( x-xy-y^{2}\right) \left[ \frac{1}{%
m^{2}(x+y)^{2}-q^{2}xy}-\frac{1}{m^{2}(x+y)^{2}+m_{P}^{2}\left( 1-x-y\right)
-q^{2}xy}\right] .  \label{eq 2.5}
\end{equation}%
}With these explicit expressions for: $\Pi $, $\Sigma $ and $\Lambda $, at
one-loop approximation, we now proceed to the calculation for the
counter-terms $\delta _{Z_{i}}$. These quantities are obtained through a rather lengthy
but straightforward calculation, only requiring that the renormalization conditions be satisfied.

\subsection{Renormalization constant $Z_{3}$}

We begin from the simplest counter-term calculation; which is from
the gauge sector. From the relations \eqref{eq 1.10} and \eqref{eq 2.2} one can easily find:
\begin{equation}
\delta _{Z_{3}}^{\left( 1\right) }=\left( Z_{3}-1\right) ^{\left( 1\right) }=%
\frac{\alpha }{3\pi }\left[ \frac{2}{\epsilon _{UV}}-\frac{1+2\gamma }{2}%
+\ln \left[ \frac{4\pi \mu ^{2}}{m^{2}}\right] \right] .  \label{eq 2.6}
\end{equation}%
Actually, the counter-term $\delta _{Z_{3}}$ is the only one which has an ultraviolet term and is
infrared divergence free.

\subsection{Renormalization constant $Z_{0}$}

In the fermionic sector, we first compute the counter-term
related with the massive fermionic sector by calculating the Eq.\eqref{eq 1.16}. One can easily cast the Eq.\eqref{eq
2.2} in the form $\Sigma \left( p,m\right) =\Sigma _{1}\left(
p^{2}\right) \widehat{p}+\Sigma _{2}\left( p^{2}\right) I$. Therefore, from
Eq.\eqref{eq 1.16}, the resulting expression is:
\begin{eqnarray}
\delta _{Z_{0}}^{\left( 1\right) }=\frac{\alpha }{2\pi }
\frac{\left( 3-\xi \right)}{\epsilon _{IR}} +\frac{\alpha }{4\pi }\left( 1+4b -\left( 2b^{2}+\xi
-3\right) \log (b)\right)  +\frac{\alpha }{2\pi }\frac{b((b-2)b-5)}{\sqrt{b(b-4)}}\ln \left[
\frac{b+\sqrt{b\left( b-4\right) }}{b-\sqrt{b\left( b-4\right) }}\right] .\label{eq 2.14}
\end{eqnarray}
Where we have introduced the infrared dimensional parameter: $\epsilon _{IR}=d-4$, $\epsilon _{IR}\rightarrow 0^{-}$. It was considered the region: $b=\frac{m_{P}^{2}}{m^{2}}>4$. Otherwise,
the logarithm must be replaced by an $\arctan $ function in the above
expression. We have here obtained an ultraviolet finite counter-term, which
apparently might indicate a finite renormalization constant; however, it is, in fact, infrared divergent at order-$\alpha $. Anyhow, we will present a proper discussion regarding this issue right below.

\subsection{Renormalization constant $Z_{2}$ and $Z_{1}$}

The counter-term $\delta _{Z_{2}}$ can be obtained from Eq.\eqref{eq 1.13c}. Hence, follows that, at order-$\alpha $, it has the following expression for $b>4$:{\small%
\begin{eqnarray}
\delta _{Z_{2}}^{\left( 1\right) } &=&\frac{\alpha }{2\pi }
\frac{\left( 3-\xi \right)}{\epsilon _{IR}}+\frac{\alpha }{24\pi }\left(36+18b-6\xi+ 12b^{2}\xi\right) + \frac{\alpha }{8\pi }\left( -2b^{3}\xi +3b^{2}\left( \xi-1\right) +2b\xi -2\xi +6\right) \log \left( b\right) \notag \\
&&+\frac{\alpha }{8\pi }\frac{b}{\sqrt{b\left( b-4\right) }}\left( 2b^{3}\xi +b^{2}\left( 3-7\xi
\right) -6b-12\right) \left[ \ln \left( \frac{b-2+\sqrt{(b-4)b}}{b-2-\sqrt{%
(b-4)b}}\right) +\ln \left( \frac{b-\sqrt{b(b-4)}}{b+\sqrt{b(b-4)}}\right) %
\right] .  \label{eq 2.11}
\end{eqnarray}
}Such expression is also an ultraviolet finite quantity at order-$\alpha $.

At last, it is only remaining the vertex renormalization constant $%
Z_{1}$ to be determined which follows from the condition \eqref{eq 1.18} and Eq.\eqref{eq 2.4}; and, after some simple integral manipulations, one finds
the same expression than Eq.\eqref{eq 2.11}. A result which is in agreement with the equality $Z_{1}=Z_{2}$ at order-$\alpha $.

Although there is a massive sector on the the propagator Eq.\eqref{eq 2.1}, the expressions of the following counter-terms: $\delta _{Z_{0}}$ Eq.\eqref{eq 2.14}, $\delta _{Z_{2}}$ and $\delta _{Z_{1}}$ Eq.$\left( \ref{eq 2.11}\right) $, reveal that all of them have an infrared divergence which comes from the usual massless $QED_{4}$ sector contribution. Nevertheless, it possesses a simple $\xi$-dependent structure which is easily recognized as being:
\begin{equation}
I_{IR}=\frac{\alpha }{2\pi }\frac{\left( 3-\xi \right)}{\epsilon _{IR}}.
\label{eq 2.17}
\end{equation}%
By a renormalization group analysis in $QED_{4}$%
, in $p^{2}\ll m^{2}$ regime, it is known that there is unique gauge
choice in which the electron propagator behaves asymptotically free and it
is also infrared safe; this gauge choice is known as Fried-Yennie gauge \cite%
{16}, and stands as $\xi =3$. The same statement holds here to the $GQED_{4}$
(see Eq.\eqref{eq 2.17}). Therefore, as it turns out from one-loop
calculation, we can conclude that the choice $\xi =3$, besides the $\Pi $ expression, yields to
ultraviolet and infrared finite expressions for $GQED_{4}$.


\section{$GQED_{4}$ Effective Coupling}

\label{sec:3}

As it has been mentioned earlier, although the renormalization constant $Z_{3}$ Eq.\eqref{eq
2.6} does not depend on Podolsky's parameter $m_{P}$, it might there to be
another quantities that may be sensitive to these effects in order-$\alpha $. A suitable scenario to investigate such effects would be the Coulomb scattering \cite{17}. In fact, the Coulomb potential of a point-like charge is changed in Generalized Electrodynamics as: $\phi(r)=\frac{e}{r}%
\left(1-e^{-m_{P}r}\right)$. A proper way of representing an important class of these modifications is to
introduce some invariant quantities, such as the so-called running coupling constant. Therefore, it is the purpose of this section to examine the effects of
Podolsky Electrodynamics into the usual running coupling constant in the Coulomb scattering.

In fact, we are interested in modifications of the Coulomb scattering
amplitude at $k^{2}/m^{2}\gg 1$ regime. In this approximation, one defines the running coupling constant as follows:{\small %
\begin{eqnarray*}
&& \alpha _{R}\left( k^{2}\right) =\alpha \Bigg\{ 1+\left( 1-\frac{k^{2}}{%
k^{2}-m_{P}^{2}}\right) \left[ \frac{\alpha }{3\pi }\left[ \frac{2}{\epsilon
}-\frac{1+2\gamma }{2}+\ln \left[ \frac{4\pi \mu ^{2}}{m^{2}}\right] \right]
+\frac{\alpha }{3\pi }\ln \left( \frac{k^{2}}{m^{2}}\right) \right] \Bigg\},
\end{eqnarray*}
}for which one immediately gets:{\small %
\begin{eqnarray}
\alpha _{R}\left( k^{2}\right) =\alpha _{R}\left( m^{2}\right) \Bigg[ 1+%
\frac{\alpha _{R}\left( m^{2}\right) }{3\pi }\frac{1}{1-\frac{k^{2}}{%
m_{P}^{2}}}\ln \left[ \frac{k^{2}}{m^{2}}\right]  \Bigg] ,  \label{eq 3.5}
\end{eqnarray}%
}where $\alpha _{R}\left( m^{2}\right) =Z_{3}\alpha $, with $Z_{3}$ given by
Eq.\eqref{eq 2.6}.

Furthermore, the modifications to the Coulomb scattering can also be obtained
in higher-orders of perturbation theory in the regime $k^{2}/m^{2}\gg
1$. For this purpose, we can sum an important class of diagrams through the
leading logarithmic approximation; which corresponds to the most divergent
set of logarithms. Hence, by means of this approximation we can cast the running coupling constant expression into the following form:
\begin{equation}
\frac{1}{\alpha _{R}\left( k^{2}\right) }=\frac{1}{\alpha _{R}\left(
m^{2}\right) }-\frac{1}{3\pi }\frac{1}{1-\frac{k^{2}}{m_{P}^{2}}}\ln \left[
\frac{k^{2}}{m^{2}}\right] .  \label{eq 3.7}
\end{equation}%
We see in \eqref{eq 3.7} that the running coupling constant expression clearly has a pole at $k^{2}=m_{P}^{2}$; and, in comparison to the $QED_{4}$ expression \cite{17} it provides a validity regime: $m^{2} \leq k^{2} < m_{P}^{2}$, where the $GQED_{4}$ theory is in fact well-defined. The aim of the next section is to determine a bound value for the parameter $m_{P}^{2}$, which therefore will be important to define the accessible energy regime to the theory.


\section{Electron Magnetic Moment and Podolsky's Parameter}

\label{sec:4}

The electron anomalous magnetic moment value is the most accurate
test of particle physics up-to-date. It was calculated initially at one-loop
by Schwinger \cite{18}, and at two-loop in \cite{19,21}. The subsequent orders
calculation was summarized by Kinoshita et al \cite{20}.

We have as purpose in this section, based on experimental grounds, to make use of this precise
data to set a bound value to the Podolsky's parameter $m_{p}$ in high-energy
physics. Thereafter, we shall proceed into the calculation of the Pauli's form factor
$F_{2}\left( q^{2}\right) $, Eq.\eqref{eq 2.5}, which can be
written as the sum of two contributions:%
\begin{equation}
F_{2}\left( q^{2}\right) =F_{QED}\left( q^{2}\right) +F_{POD}\left(
q^{2}\right) .  \label{eq 4.1}
\end{equation}%
The first term, $F_{QED}\left( q^{2}\right) $, is the well-known one-loop
contribution calculated by Schwinger \cite{18}, and it reads:
\begin{equation}
F_{QED}\left( 0\right) =\frac{\alpha }{2\pi }.  \label{eq 4.2}
\end{equation}%
Whereas the second term $F_{POD}\left( q^{2}\right) $ gives a new and
interesting contribution. We have to solve here the following integral:{\small%
\begin{equation}
F_{POD}\left( 0\right) =\frac{\alpha }{\pi }\int_{0}^{1}du\frac{u^{3}-u^{2}}{%
u^{2}+\frac{m_{P}^{2}}{m^{2}}(1-u)},  \label{eq 4.3}
\end{equation}%
}Hence, assuming the condition $b>4$, with $b=\frac{m_{P}^{2}}{m^{2}}$, one obtains:{\small%
\begin{eqnarray}
F_{POD}\left( 0\right) =\frac{\alpha }{2\pi }\Bigg[ 2b-1+\left(2-b\right) b\ln \left(b\right)  -\frac{b\left( 2+b\left( b-4\right)\right) }{\sqrt{b\left(b-4\right) }} \left\{ \ln \left( \frac{\sqrt{%
b\left( b-4\right) }+2-b}{\sqrt{b\left( b-4\right) }-2+b}\right) +\ln \left(
\frac{\sqrt{b\left( b-4\right) }+b}{\sqrt{b\left( b-4\right) }-b}\right)
\right\} \Bigg].  \label{eq 4.5}
\end{eqnarray}%
}
We claim here to state that, since we have a perfect agreement of
the $QED_{4}$ results with experiments, the Podoslky contribution Eq.\eqref{eq 4.5} must be at most equal to the experimental
error in the electron anomalous magnetic moment value. The experimental
value of electron anomalous magnetic moment is $a_{\exp}=1,15965218073\times 10^{-3}\pm 2,8\times 10^{-13}$ \cite{22}. Therefore, from the expression \eqref{eq 4.5} we find that for values: $m_{P}\geq3,7595\times 10^{10}eV$ the theory is compatible with the
experimental data. \footnote{Actually, we find a consistent value for $m_{P}$ in the region $b>4$ only.}


\section{Concluding Remarks}


\label{sec:5}

This paper presents a proper study regarding the renormalizability, and subsequent consequences,
of the Generalized Quantum Electrodynamics. It was evoked initially arguments about the consistency of Generalized Electrodynamics, the construction of the physical subspace through a couple of possible prescriptions, where it is always claimed the presence of a powerful symmetry: BRST symmetry, reflection positivity and etc.

We had successfully applied the renormalization program on $GQED_{4}$, and subsequent quantities were computed at one-loop approximation. Once we had an UV finite theory \cite{11} and also identified the primitive divergences, it was suitable to examine the theory at renormalizability level, what took us to develop the renormalization prescription for the Lagrangian density. Although we had chosen a condition where the photon propagator was massless, there were some quantities affected by the theory's free parameter $m_{P}$.

Next we presented the expressions for all four counter-terms of theory. Although the fermionic and vertex
counter-terms: $\delta _{Z_{0}}$, $\delta _{Z_{2}}$, $\delta_{Z_{1}}$, have
ultraviolet finite expressions, an analysis allowed us to identify that all expressions had the same infrared
term, introduced by the integration over the Feynman parameters. However,
such infrared term had a simple dependence in the gauge parameter $\xi $ as: $\left(
3-\xi \right) $, which is similar to the Fried-Yennie gauge, $\xi =3$.
This gauge choice is well-known as being infrared safe in $QED_{4}$.
The same conclusion occurs in $GQED_{4}$.

Although the photon field does not feel the effects from the Podolsky contribution
at order-$\alpha$, we studied the scenario of the Coulomb
scattering, which actually is sensitive to these effects. Actually, the invariant quantity studied here was the effective coupling of the theory, i.e., the running coupling constant. Hence, through its explicit and $m_{P}$-dependent expression, we were able to find a energy range of validity for the $GQED_{4}$: $m^{2} \leq k^{2} < m_{P}^{2}$.

The last part of work was addressed to the discussion and evaluation
of the Podolsky contribution to the electron anomalous magnetic moment at
one-loop approximation. Firstly we obtained the known result, the $QED_{4}$
contribution: $F_{2}\left( 0\right) =\frac{\alpha }{2\pi }$. Next, we proposed the possibility of using the
experimental data of the electron anomalous magnetic moment to set a limit to
the Podolsky parameter $m_{P}$. Therefore, we found a
consistent value for this parameter as being: $m_{P}\geq3,7595\times 10^{10}eV$. It is worth to stress here that, as the theory has a natural energy scale, which is the electron mass $m$, it provides a sensitive result to the
Podolsky parameter value.

There are numerous extensions one may consider. For instance, owing its rich interacting structure, the Scalar Generalized Electrodynamics and its non-abelian version would be an appealing extensions, once we now have good insights about the behavior of the complete theory. A study of HD of gravitational fields should also be interesting, it continues being an open issue, and revisited in several different approaches and in different dimensionality. Still in the $GQED_{4}$ context, it might be an interesting case the study of scattering process with external fields of the HD theories \cite{23}. However, we believe that the remaining major issue involving $GQED_{4}$, is to formulate the theory following the causal perturbation theory \cite{24};
which by itself is well-established, leading to a well-defined and ultraviolet finite theory. These issues and others will be further
elaborated, investigated and reported elsewhere.


\subsection*{Acknowledgements}


The authors would like to thank the referee for his/her comments and suggestions to make this paper clearer. RB thanks CNPq and FAPESP for full support, BMP thanks CNPq and CAPES for partial support and GERZ thanks VIPRIUDENAR for partial support.

\end{document}